\documentclass[a4paper]{jpconf}
\usepackage{graphicx}
\usepackage{amsmath}
\usepackage{amssymb}

\begin{document}
\title{Vector mesons spectrum in a medium with a chiral imbalance induced by the vacuum of fermions}

\author{Vladimir Kovalenko, Alexander Andrianov and Vladimir Andrianov}

\address{Saint Petersburg State University, 199034,  Universitetskaya nab. 7/9, St. Petersburg, Russia}

\ead{v.kovalenko@spbu.ru}

\begin{abstract}
The properties of the light vector mesons in the presence of local parity breaking medium with a chiral imbalance are considered in the  vector-meson dominance model. Applying the finite lowest-order radiatively induced local effective Lagrangian, initially developed for the QED, for the vector $\rho$ and $\omega$ mesons, we obtained the mass spectrum as a function of momentum and chiral chemical potential $\mu_5$. We showed that in addition to the Chern–Simons term, splitting the transverse polarisations of the mesons, there is another radiatively induced contribution that becomes important at momentum and $\mu_5$ around a few hundred MeV.
\end{abstract}

\section{Introduction}
The chiral imbalance defined as an average difference between the numbers of RH and LH quarks may occur in the fireball after a heavy-ion collision at high energy. It can lead to the formation of a local parity breaking (LPB) in a quark-hadron medium and be adiabatically characterized by a chiral chemical potential $\mu_5$.  The consistent way for the construction of the Lagrangian in the presence of constant axial-vector background has been obtained in [1, 2] for QED theory.

The general idea of this work is to make use the finite lowest-order radiatively induced local effective Lagrangian (\ref{LIVgamma}) \cite{Alfaro:2006dd,Alfaro:2009mr} for the LIV $\tilde{\gamma}$-photon, and apply it for vector meson field containing the lightest
vector mesons $\rho_0$ and $\omega$ in the SU(2) flavor sector.

\section{Effective LIV Lagranginan for VDM fields}
Schematically the effective lagrangian for vector fields including radiative 1-loop contributions takes the form
\begin{equation}\label{LIVgamma}
\begin{aligned}
\mathcal{L}_{\mathrm{eff}}=-& \frac{1}{4} F^{\mu \nu} F_{\mu \nu}\left(1+\frac{2 \alpha b^{2}}{3 \pi m^{2}}\right)+\frac{\alpha b^{2}}{3 \pi} A^{\mu} A_{\mu} \\
&+\frac{\alpha}{3 \pi m^{2}} b_{\nu} b^{\rho} F^{\nu \lambda} F_{\rho \lambda}-\frac{\alpha}{2 \pi} b_{\lambda} A_{\mu} \epsilon^{\lambda \mu \rho \sigma} F_{\rho \sigma} \\
=&-\frac{1}{4}(1+\varepsilon) F^{\mu \nu} F_{\mu \nu}+\frac{1}{2} \varepsilon m^{2} A^{\mu} A_{\mu} \\
&+\varepsilon \frac{b^{\lambda} b^{\nu}}{2 b^{2}} F_{\lambda \rho} F_{\nu}^{\rho}-\frac{\alpha}{\pi} b_{\mu} A_{\nu} \tilde{F}^{\mu \nu}
 \quad\left(\varepsilon=\frac{2 \alpha b^{2}}{3 \pi m^{2}}\right)
\end{aligned}
\end{equation}
Here $m$ is a constituent quark mass ($b^2 < m^2$).
\clearpage

The VDM Lagrangian in the SU(2) flavor sector reads [3]:
\begin{equation}
\mathcal{L}_{\text {int }}=\bar{q} \gamma_{\mu} V^{\mu} q 
\end{equation}
\begin{equation}\label{VectorField}
V_{\mu} \equiv-e A_{\mu} Q+\frac{1}{2} g_{\omega} \omega_{\mu} \mathbf{I}_{q}+\frac{1}{2} g_{\rho} \rho_{\mu} \lambda_{3}
\end{equation}

where $Q=\frac{\lambda_{3}}{2}+\frac{1}{6} \mathbf{I}_{q}, g_{\omega} \simeq g_{\rho} \equiv g \simeq 6; \mathbf{I}_{q}$ is the identity matrix in isospin space and $\lambda_{3}$ is the corresponding Gell-Mann matrix. 
The Maxwell and mass terms are
\begin{equation}\label{kin}
\mathcal{L}_{\mathrm{kin}}=-\frac{1}{4}\left(F_{\mu \nu} F^{\mu \nu}+\omega_{\mu \nu} \omega^{\mu \nu}+\rho_{\mu \nu} \rho^{\mu \nu}\right)+\frac{1}{2} V_{\mu, a} m_{a b}^{2} V_{b}^{\mu}
\end{equation}
\begin{equation}\label{massmatrix}
m_{a b}^{2}=m_V^{2}\left(\begin{array}{ccc}\frac{10 e^{2}}{9 g^{2}} & -\frac{e}{3 g} & -\frac{e}{g} \\
-\frac{e}{3 g} & 1 & 0  \\ 
-\frac{e}{g} & 0 & 1  \\
\end{array}\right), \quad \operatorname{det}\left(m_{a b}^{2}\right)=0
\end{equation}

where $\left(V_{\mu, a}\right) \equiv\left(A_{\mu}, \omega_{\mu}, \rho_{\mu}^{0} \equiv \rho_{\mu}\right)$ and $m_V^{2}=m_{\rho}^{2},  g_{\rho}\simeq g_{\omega} \equiv g \simeq 6$  \cite{Andrianov:2012hq}.

Let us replace in (\ref{LIVgamma}) $A_\mu$ by $V_\mu$ defined as in (\ref{VectorField}). Doing such substitution we must wipe the coupling constants in (\ref{LIVgamma}) and replace the kinetic term by (\ref{kin}). We obtain the following effective Lagrangian:
\begin{equation}
\mathcal{L}_{\mathrm{eff}} = \mathcal{L}_{\mathrm{kin}} 
-\frac{1}{4} \varepsilon V_{\mu\nu} \varepsilon V^{\mu\nu} + \frac{1}{2}\varepsilon m^2 V_\nu V^\nu + \varepsilon \frac{b^{\lambda} b^{\nu}}{2 b^{2}} V_{\lambda \rho} V_{\nu}^{\rho}-\frac{N_c}{8\pi^2} b_{\mu} V_{\nu} \epsilon^{\mu \nu \lambda \nu} V_{\lambda \nu},
\end{equation}
where 
\begin{equation}
\varepsilon=\frac{b^2 N_c}{6\pi^2 m} ^\text{\footnotemark} \text{ and } V_{\mu\nu}=\partial_\mu V_\nu - \partial_\nu V_\mu
\end{equation}
\footnotetext{in (\ref{LIVgamma}) we replaced $\alpha$ with $\frac{g^2}{4\pi}$ and then removed $g^2$.}
Trace over isospin matrices is assumed.

For the massive fremion mass we use $m$ as a dynamical quark mass.

As an intermediate result, 

\begin{equation}
\begin{aligned}
\mathcal{L}_{\mathrm{eff}}=
-\frac{F_{\mu v} F^{\mu v}}{4}
-\frac{\omega_{\mu v} \omega^{\mu v}}{4}
-\frac{\rho_{\mu v} \rho^{\mu v}}{4}\hspace{4.6cm}\\
+\frac{5 m_V^{2} e^{2} A_{\mu} A^{\mu}}{9 g^{2}}-\frac{ m_V^{2} e A^{\mu} \omega_{\mu}}{3 g}-\frac{ m_V^{2} e A^{\mu} \rho_{\mu}}{g}+\frac{b^2 N_c V^{v} V_{v}}{24 \pi^{2} m^{2}} \\
-\frac{N_c \epsilon_{\delta \gamma \mu v} b^{\mu} V^{v} V^{\delta \gamma}}{8 \pi^{2}}+\frac{V_{\gamma \lambda} N_c V_{\mu}^{\lambda} b^{\gamma} b^{\mu}}{12 \pi^{2} m^{2}}+\frac{1}{2}m_V^2 \omega_\mu \omega^\mu+\frac{1}{2} m_V^{2} \rho_{\mu} \rho^{\mu}
\end{aligned}
\end{equation}

Let us apply the definition of a vector meson field (\ref{VectorField}).
Then, after applying the trace summation over the isospin matrices\footnote{We used $\text{Tr} \mathbf{I}_{q}^2 = 2, \text{ Tr} \lambda_3^2 = 2, \text{Tr} \mathbf{I}_{q} \lambda_3 = 0.$}, we obtain:

\begin{equation}\begin{aligned}
&\mathcal{L}_{\mathrm{eff}}=-\frac{F_{\mu \nu} F^{\mu \nu}}{4}-\frac{\omega_{\mu \nu} \omega^{\mu \nu}}{4}-\frac{\rho_{\mu \nu} \rho^{\mu \nu}}{4} +\frac{1}{2}m_V^{2} \omega_{\mu} \omega^{\mu}+\frac{1}{2} m_V^{2} \rho_{\mu} \rho^{\mu}\\
&+\frac{b^2 N_c e F_{\mu \nu} g \omega^{\mu \nu}}{144 \pi^{2} m^{2}}+\frac{N_c \epsilon_{\mu \nu \gamma \delta} b^{\mu} \rho^{\nu} g e F^{\gamma \delta}}{16 \pi^{2}}+\frac{N_c b^{\lambda} b^{\nu} g^{2} \omega_{\lambda \sigma} \omega_{\nu}^{\sigma}}{24 \pi^{2} m^{2}}\\
&+\frac{b^2 N_c e F_{\mu \nu} g \rho^{\mu \nu}}{48 \pi^{2} m^{2}}+\frac{b^2 N_c g \rho_{\mu \nu} e F^{\mu \nu}}{48 \pi^{2} m^{2}}+\frac{N_c \epsilon_{\mu \nu \gamma \delta} b^{\mu} e A^{\nu} g \omega^{\gamma \delta}}{48 \pi^{2}}\\
&+\frac{5 N_c b^{\lambda} b^{\nu} e^{2} F_{\lambda \sigma} F_{\nu}^{\sigma}}{108 \pi^{2} m^{2}}+\frac{N_c b^{\lambda} b^{\nu} g^{2} \rho_{\lambda \sigma} \rho_{\nu}^{\sigma}}{24 \pi^{2} m^{2}}+\frac{N_c \epsilon_{\mu \nu \gamma \delta} b^{\mu} e A^{\nu} g \rho^{\gamma \delta}}{16 \pi^{2}}\\
&+\frac{N_c \epsilon_{\mu \nu \gamma \delta} b^{\mu} \omega^{\nu} g e F^{\gamma \delta}}{48 \pi^{2}}+\frac{b^2 N_c g \omega_{\mu \nu} e F^{\mu \nu}}{144 \pi^{2} m^{2}}-\frac{5 b^2 N_c e^{2} F_{\mu \nu} F^{\mu \nu}}{216 \pi^{2} m^{2}}\\
&-\frac{N_c \epsilon_{\mu \nu \gamma \delta} b^{\mu} \omega^{\nu} g^{2} \omega^{\gamma \delta}}{16 \pi^{2}}-\frac{b^2 N_c \omega^{\nu} g e A_{\nu}}{72 \pi^{2}}-\frac{5 N_c \epsilon_{\mu \nu \gamma \delta} b^{\mu} e^{2} A^{\nu} F^{\gamma \delta}}{72 \pi^{2}}-\frac{b^2 N_c e A^{\nu} \omega_{\nu} g}{72 \pi^{2}}\\
&-\frac{b^2 N_c g^{2} \omega_{\mu \nu} \omega^{\mu \nu}}{48 \pi^{2} m^{2}}-\frac{N_c \epsilon_{\mu \nu \gamma \delta} b^{\mu} \rho^{\nu} g^{2} \rho^{\gamma \delta}}{16 \pi^{2}}-\frac{b^2 N_c g^{2} \rho_{\mu \nu} \rho^{\mu \nu}}{48 \pi^{2} m^{2}}-\frac{b^2 N_c \rho^{\nu} g e A_{\nu}}{24 \pi^{2}}\\
&-\frac{b^2 N_c e A^{\nu} \rho_{\nu} g}{24 \pi^{2}}+\frac{5 m_V^{2} e^{2} A_{\mu} A^{\mu}}{9 g^{2}}-\frac{m_V^{2} e A_{\mu} \omega^{\mu}}{6 g}-\frac{m_V^{2} e A_{\mu} \rho^{\mu}}{2g}-\frac{m_V^{2} e A^{\mu} \omega_{\mu}}{6 g}-\frac{m_V^{2} e A^{\mu} \rho_{\mu}}{2g}\\
&+\frac{5 b^2 N_c e^{2} A^{\nu} A_{\nu}}{108 \pi^{2}}+\frac{b^2 N_c \omega^{\nu} g^{2} \omega_{\nu}}{24 \pi^{2}}+\frac{b^2 N_c \rho^{\nu} g^{2} \rho_{\nu}}{24 \pi^{2}}-\frac{N_c b^{\lambda} b^{\nu} g \rho_{\lambda \sigma} e F_{\nu}^{\sigma}}{24 \pi^{2} m^{2}}\\
&-\frac{N_c b^{\lambda} b^{\nu} e F_{\lambda \sigma} g \omega_{\nu}^{\sigma}}{72 \pi^{2} m^{2}}-\frac{N_c b^{\lambda} b^{\nu} g \omega_{\lambda \sigma} e F_{\nu}^{\sigma}}{72 \pi^{2} m^{2}}-\frac{N_c b^{\lambda} b^{\nu} e F_{\lambda \sigma} g \rho_{\nu}^{\sigma}}{24 \pi^{2} m^{2}}
\end{aligned}\end{equation}

In order to allow the physical treatment, we will apply an orthogonal transformation in the space of three vector mesons. The goal is to diagonalize the vector meson mass quadratic from. The LIV-terms will also be diagonalized. The kinetic term keeps invariant. 

The mass matrix is defined in (\ref{massmatrix}).
It's eignvalues are $0,\left(1+\dfrac{10 e^2}{9 g^2}\right) m_V,m_V$ correspond to masses of photon, $\omega$ and $\rho$ mesons.

The transform matrix, built from orthonormalized eignvectors, reads:

\begin{equation}U=
\left(
\begin{array}{ccc}
\frac{3 g}{\sqrt{\frac{9 g^{2}}{e^{2}}+10} e} & -\frac{10 e}{\sqrt{\frac{100 e^{2}}{g^{2}}+90} g} & 0 \\
\frac{1}{\sqrt{\frac{9 g^{2}}{e^{2}}+10}} & \frac{3}{\sqrt{\frac{100 e^{2}}{g^{2}}+90}} & -\frac{3 \sqrt{10}}{10} \\
\frac{3}{\sqrt{\frac{9 g^{2}}{e^{2}}+10}} & \frac{9}{\sqrt{\frac{100 e^{2}}{g^{2}}+90}}  & \frac{\sqrt{10}}{10}
\end{array}
\right)
\end{equation}

We define a new vector field $V'^{\,\mu}$ so that $V^{\ \mu}_a = U_{ab} V'^{\ \mu}_b $.

Then, after simplifications and removing primes from the new vector field, we obtain the following equation for the effective Lagrangian:

\begin{equation}\label{ourresult}\begin{array}{l}
\mathcal{L}_{\mathrm{eff}}=-\frac{F_{\mu \nu} F^{\mu \nu}}{4}+\left(-\frac{1}{4}-\frac{5 b^{2} N_{c} e^{2}}{216 \pi^{2} m^{2}}-\frac{g^{2} b^{2} N_{c}}{48 \pi^{2} m^{2}}\right) \omega^{\delta \gamma} \omega_{\delta \gamma}
+\left(-\frac{1}{4}-\frac{g^{2} b^{2} N_{c}}{48 \pi^{2} m^{2}}\right) \rho^{\delta \gamma} \rho_{\delta \gamma}\\
+\left(\frac{1}{2}m_V^{2}+\frac{5 e^{2} m_V^{2}}{9 g^{2}}+\frac{5 N_{c} b^{2} e^{2}}{108 \pi^{2}}+\frac{g^{2} b^{2} N_{c}}{24 \pi^{2}}\right) \omega^{\mu} \omega_{\mu}+\left(\frac{1}{2}m_V^{2}+\frac{g^{2} b^{2} N_{c}}{24 \pi^{2}}\right) \rho^{\mu} \rho_{\mu}\\
+\left(\frac{5 N_{c} b^{\delta} b^{\lambda} e^{2}}{108 \pi^{2} m^{2}}+\frac{g^{2} N_{c} b^{\delta} b^{\lambda}}{24 \pi^{2} m^{2}}\right) \omega_{\delta \gamma} \omega_{\lambda}^{\gamma}+\frac{g^{2} N_{c} b^{\delta} b^{\lambda} \rho_{\delta \gamma} \rho_{\lambda}^{\gamma}}{24 \pi^{2} m^{2}} \\
+\left(-\frac{5 N_{c} e^{2} b^{\lambda} \epsilon_{\delta \gamma \lambda \mu}}{72 \pi^{2}}-\frac{g^{2} N_{c} \epsilon_{\delta \gamma \lambda \mu} b^{\lambda}}{16 \pi^{2}}\right) \omega^{\mu} \omega^{\delta \gamma} 
-\frac{g^{2} N_{c} \epsilon_{\delta \gamma \lambda \mu} b^{\lambda} \rho^{\mu} \rho^{\delta \gamma}}{16 \pi^{2}}
\end{array}\end{equation}

It is clear that the electromagnetic field completely decouples as a free massless field. The vector mesons decouple between each other. 

Let us write the vector meson Lagrangian part (\ref{ourresult}) in the form similar to \cite{Alfaro:2006dd}:

\begin{equation}\label{vectorpart}\begin{array}{l}
\mathcal{L}_{V}=
-\dfrac{1}{4}\left(1+\xi \dfrac{b^2}{m^2} \right)V^{\mu\nu} V_{\mu\nu}
+\xi \dfrac{ b_\nu b^\rho }{2 m^2} V^{\nu\lambda} V_{\rho\lambda}\\
-\dfrac{1}{2}\zeta b_\nu V_\lambda \dfrac{1}{2} \epsilon^{\nu\lambda\rho\sigma}V_{\rho\sigma} + \dfrac{1}{2}\bar{m}^2 V_\nu V^\nu
\end{array}\end{equation}
where
\begin{equation}
\begin{array}{c}
\xi=\left\lbrace 
\begin{array}{l}
\xi_\omega =  \dfrac{ g^2 N_c }{12 \pi^2 } 
           + \dfrac{ 5 e^2 N_c } {54 \pi^2} \\
\xi_\rho = \dfrac{ g^2 N_c }{12 \pi^2 }
\end{array}
\right.
\zeta=\left\lbrace 
\begin{array}{l}
\zeta_\omega =  \dfrac{ g^2 N_c }{4 \pi^2 } 
           + \dfrac{ 5 e^2 N_c } {18 \pi^2} \\
\zeta_\rho = \dfrac{ g^2 N_c }{4 \pi^2 }
\end{array}
\right.
\\
\bar{m}^2=\left\lbrace 
\begin{array}{l}
\bar{m}_\omega^2 =  m_V^2 + \dfrac{10 e^2}{9 g^2} m_V^2 + \dfrac{ g^2 N_c }{12 \pi^2 } b^2 
 + \dfrac{5 N_c e^2}{54} b^2
 \\
\bar{m}_\rho^2 = m_V^2 + \dfrac{ g^2 N_c }{12 \pi^2 } b^2
\end{array}
\right.
\end{array}
\end{equation}

One can easily see that $\zeta=3\xi$ and
$\bar{m}(b)=\bar{m}(0)+\xi b^2$

Comparing formula (\ref{vectorpart}) with formula (9) of \cite{Andrianov:2012hq} we recognize the Chern-Simons term.

\section{Dispersion relations for $\rho$ and $\omega$}

The modified Maxwell's equations read \cite{Alfaro:2006dd}:

\begin{equation}\begin{array}{l}
\left(1+\xi \frac{b^{2}}{m^{2}}\right) \partial_{\lambda} V^{\lambda \nu}-\frac{\xi}{m^{2}}\left(b^{\rho} b_{\lambda} \partial_{\rho} V^{\lambda \nu}-b^{\nu} b_{\lambda} \partial_{\rho} V^{\lambda \rho}\right) 
+\bar{m}^{2} V^{\nu}-\zeta b_{\lambda} \tilde{V}^{\nu \lambda}=0 \\
\partial_{\nu} V^{\nu}=0
\end{array}\end{equation}

One can rewrite the field equations in
terms of $V_\nu$ only:

\begin{equation}\begin{array}{l}
\left(1+\xi \frac{b^{2}}{m^{2}}\right) \partial^{2} V^{\nu} 
-\left(\xi / m^{2}\right)\left[(b \cdot \partial)^{2} V^{\nu}-\partial^{\nu}(b \cdot \partial)(b \cdot V)+b^{\nu} \partial^{2}(b \cdot V)\right]  \\
+\bar{m}^{2} V^{\nu}-\zeta \epsilon^{\nu \lambda \rho \sigma} b_{\lambda} \partial_{\rho} V_{\sigma}=0
\end{array}\end{equation}

Multiplying by $b^\nu$ we get:

\begin{equation}\begin{array}{l}
\left(\partial^{2}+\bar{m}^{2}\right)(b \cdot V)=0
\end{array}\end{equation}

for the special component $b \cdot V$ of the vector potential.

Going to the momentum representation, the equations of motion take the form

\[
K^{\nu \sigma} A_{\sigma}(k)=0, \quad k^{\sigma} A_{\sigma}(k)=0
\]

where we have set

\[
\begin{aligned}
K^{\nu \sigma} \equiv &\left(k^{2}-\bar{m}^{2}\right) g^{\nu \sigma}-k^{\nu} k^{\sigma}-\xi\left(\mathrm{D} / m^{2}\right) \mathrm{e}^{\nu \sigma}+i \zeta \epsilon^{\nu \lambda \rho \sigma} b_{\lambda} k_{\rho}
\end{aligned}
\]

\begin{equation}\mathrm{D} \equiv(b \cdot k)^{2}-b^{2} k^{2}\end{equation}

and the projector onto the two-dimensional hyperplane orthogonal to $b_{\nu}$ and $k_{\nu}$
\begin{equation}\begin{array}{l}
\mathrm{e}^{\nu \sigma} \equiv g^{\nu \sigma}-\frac{b \cdot k}{\mathrm{D}}\left(b^{\nu} k^{\sigma}+b^{\sigma} k^{\nu}\right)+\frac{k^{2}}{\mathrm{D}} b^{\nu} b^{\sigma}+\frac{b^{2}}{\mathrm{D}} k^{\nu} k^{\sigma}
\end{array}\end{equation}

It is also convenient to define another couple of four-vectors, in order to describe the left- and right-handed polarizations: in our case, those generalize the circular polarizations of the conventional QED. To this aim, let us first define
$\epsilon^{\nu \sigma} \equiv \mathrm{D}^{-1 / 2} \epsilon^{\nu \lambda \rho \sigma} b_{\lambda} k_{\rho}$

Notice that we can always choose $\mathrm{e}_{\lambda}^{(a)}$ to satisfy
$\epsilon^{\nu \sigma} \mathrm{e}_{\sigma}^{(1)}=\mathrm{e}^{(2) \nu}, \quad \epsilon^{\nu \sigma} \mathrm{e}_{\sigma}^{(2)}=-\mathrm{e}^{(1) \nu}$

Let us now construct the two orthogonal projectors

$P_{\nu \sigma}^{(\pm)} \equiv \frac{1}{2}\left(\mathrm{e}_{\nu \sigma} \pm i \epsilon_{\nu \sigma}\right)$
\, \, and set

$\varepsilon_{\nu}^{(L)} \equiv \frac{1}{2}\left(\mathrm{e}_{\nu}^{(1)}+i \mathrm{e}_{\nu}^{(2)}\right)=P_{\nu \sigma}^{(+)} \mathrm{e}^{(1) \sigma}$

$\varepsilon_{\nu}^{(R)} \equiv \frac{1}{2}\left(\mathrm{e}_{\nu}^{(1)}-i \mathrm{e}_{\nu}^{(2)}\right)=P_{\nu \sigma}^{(-)} \mathrm{e}^{(1) \sigma}$

Then, the expression of the dispersion relations for the doubly transversal modes reads \cite{Alfaro:2006dd}:

\begin{equation}\label{dispersionrelations}\begin{array}{l}
\left\{k^{2}-\frac{\xi}{m^{2}}\left[(b \cdot k)^{2}-b^{2} k^{2}\right]-\bar{m}^{2}\right\}^{2} 
-\zeta^{2}\left[(b \cdot k)^{2}-b^{2} k^{2}\right]=0
\end{array}\end{equation}

Real solutions exist only if
\begin{equation}\nonumber
\mathrm{D}=(b \cdot k)^2-b^2 k^2 \geqslant 0    
\end{equation}

Then, for a genuine time-like $b_\mu = (b_0, 0, 0, 0)$ the Eq. (\ref{dispersionrelations}) transforms to the following:


\begin{equation}\label{dispersionrelations1}\begin{array}{l}
\left\{k_0^{2} - |\vec{k}|^2-\frac{\xi}{m^{2}}\left[b_0^2 k_0^{2}-b_0^{2} (k_0^{2} - |\vec{k}|^2)\right]-\bar{m}^{2}\right\}^{2} 
-\zeta^{2}\left[b_0^2  k_0^{2}-b_0^{2} (k_0^{2} - |\vec{k}|^2)\right]=0
\end{array}\end{equation}

The solution of the quadratic equation (\ref{dispersionrelations1}) is:

\begin{equation}\label{dispersionrelationssol}
k_0^2 - |\vec{k}|^2 \equiv {m^*}^2 = \bar{m}^2
 \pm 
 \zeta b_0 |\vec{k}|  
+ \xi b_0^2 |\vec{k}|^2/m^2 
\end{equation}

The first two terms correspond to the ones of formula (12) in \cite{Andrianov:2012hq}.

If $\bar{m}\geq \zeta b_0 /2$, then the vector meson energy
keeps real for any wave vector k and everything keeps
to be consistent. 
In our case it means:
$b_0^2 \leq 4/(5 \xi^2) m_V^2$

\section{Numerical estimations}

We use:
\begin{equation}\begin{array}{l}
N_c=3, \quad g=6, \quad e^2=4\pi \cdot 1/137, \quad 
m=0.3 \mathrm{GeV} 
 \quad m_V=m_\rho=0.7755 \mathrm{GeV}
\end{array}\end{equation}

Then

\begin{equation}
\begin{array}{c}
\xi=\left\lbrace 
\begin{array}{l}
\xi_\omega = 0.9119 + 0.0026 = 0.9145
            \\
\xi_\rho = 0.9119
\end{array}
\right.
\zeta=\left\lbrace 
\begin{array}{l}
\zeta_\omega =  2.7435\\
\zeta_\rho = 2.7357
\end{array}
\right.
\\
\bar{m}^2=\left\lbrace 
\begin{array}{l}
\bar{m}^2_\omega =  (0.7766 \mathrm{GeV})^2 + 0.9145\  b^2 
 \\
\bar{m}_\rho^2 = (0.7755 \mathrm{GeV})^2 + 0.9119 \ b^2
\end{array}
\right.
\end{array}
\end{equation}

The consistency condition for the real vector meson energy for any wave vector k is basically to $b_0 < m_V$.


The mass spectrum of vector mesons in the presence of the chiral imbalance induced by a fermionic vacuum for the transverse polarisations (\ref{dispersionrelationssol}) is:

\begin{equation}
{m_\omega^*}^2 = (0.7766 \mathrm{GeV})^2 + 0.9145\  b^2 
 \pm 
 2.7435 \ b \ |\vec{k}|  
+ 10.16 (\mathrm{GeV}^{-2}) \  b^2 \ |\vec{k}|^2
\end{equation}
\begin{equation}
{m_\rho^*}^2 = (0.7755 \mathrm{GeV})^2 + 0.9119 \ b^2 
 \pm 
 2.7357 \ b \ |\vec{k}|  
+ 10.13 (\mathrm{GeV}^{-2}) \  b^2 \ |\vec{k}|^2
\end{equation}

\begin{figure}[h!]
\includegraphics[width=0.47\textwidth]{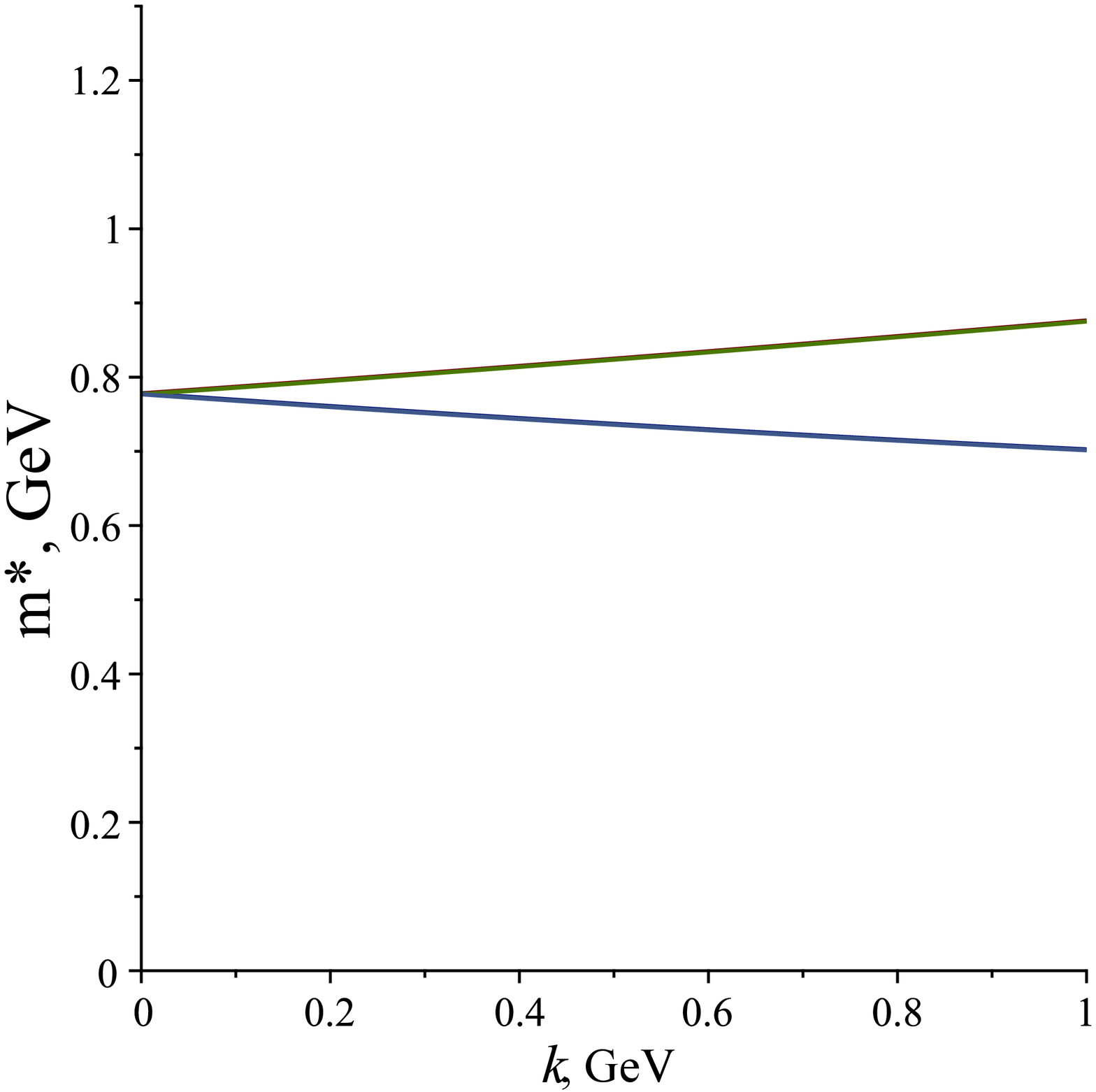} 
\includegraphics[width=0.47\textwidth]{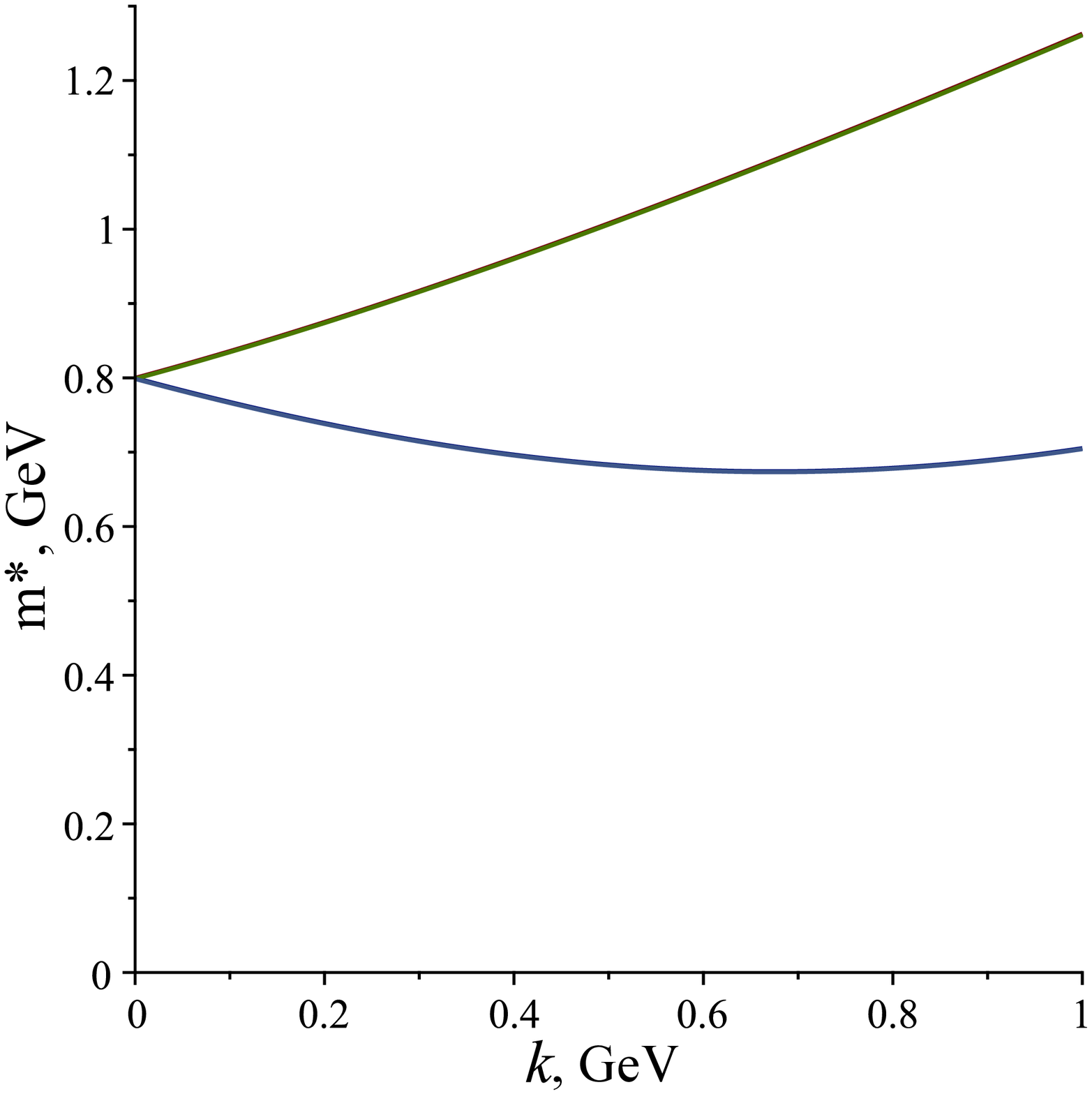} 
\vspace{-0.3cm}
\caption{Vector meson spectrum as a function of momentum at $\mu_5$=0.1 GeV and $\mu_5$=0.4 GeV}\label{fig1}
\vspace{-0.1cm}
\end{figure}

Applying the relation $\zeta b_0=N_{c} g^{2} \mu_{5} / 8 \pi^{2}$  \cite{Andrianov:2012hq}, which means $b_0 \simeq 0.5\mu_5$, we plot the mass splitting as a function of $|\vec{k}|$ at $\mu_5$=0.1 GeV, $\mu_5$=0.4 GeV (see figure \ref{fig1}). On the figure $\rho$ and $\omega$ plots overlap.

\begin{figure}[h!]
\includegraphics[width=0.47\textwidth]{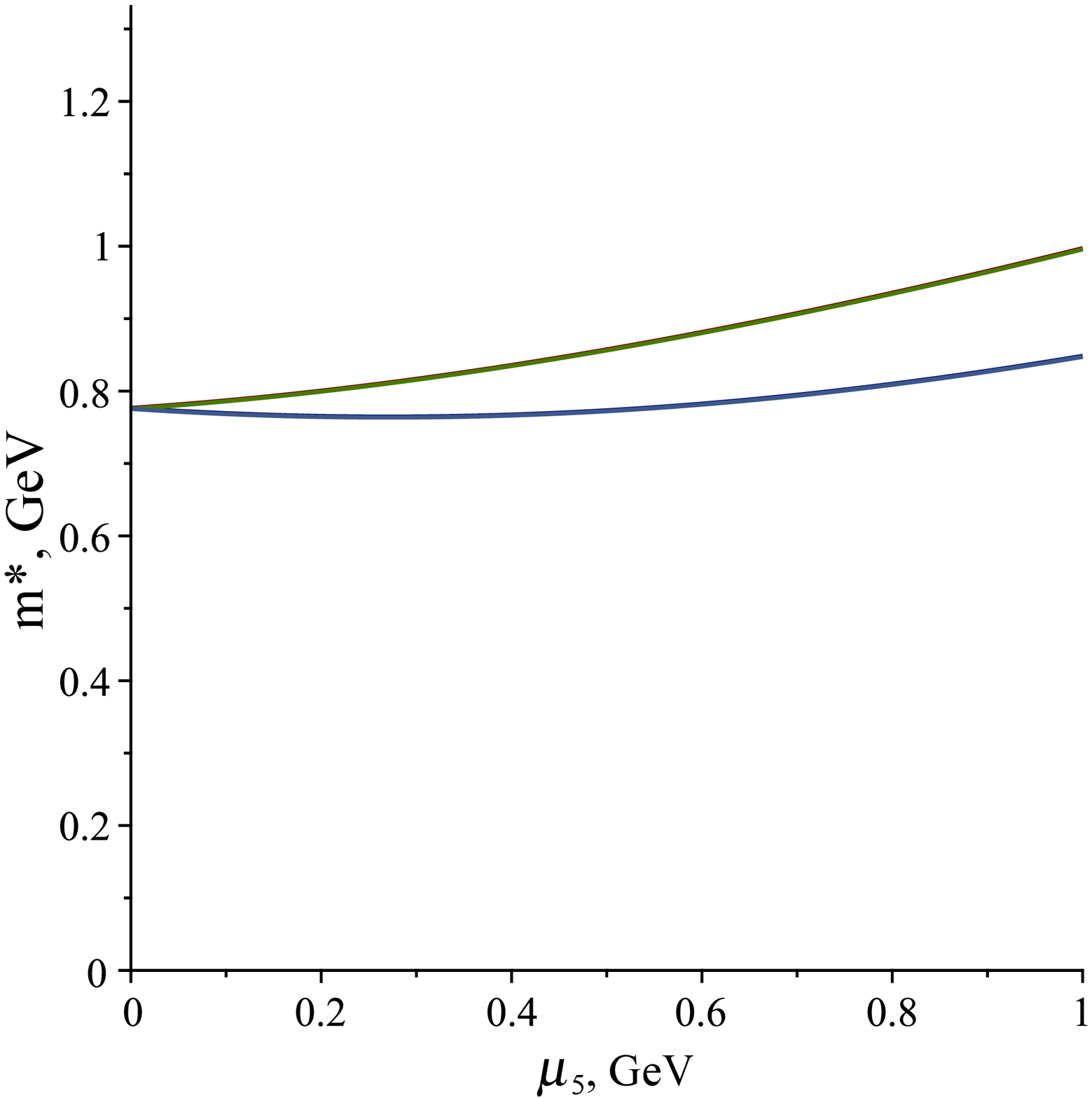} 
\includegraphics[width=0.47\textwidth]{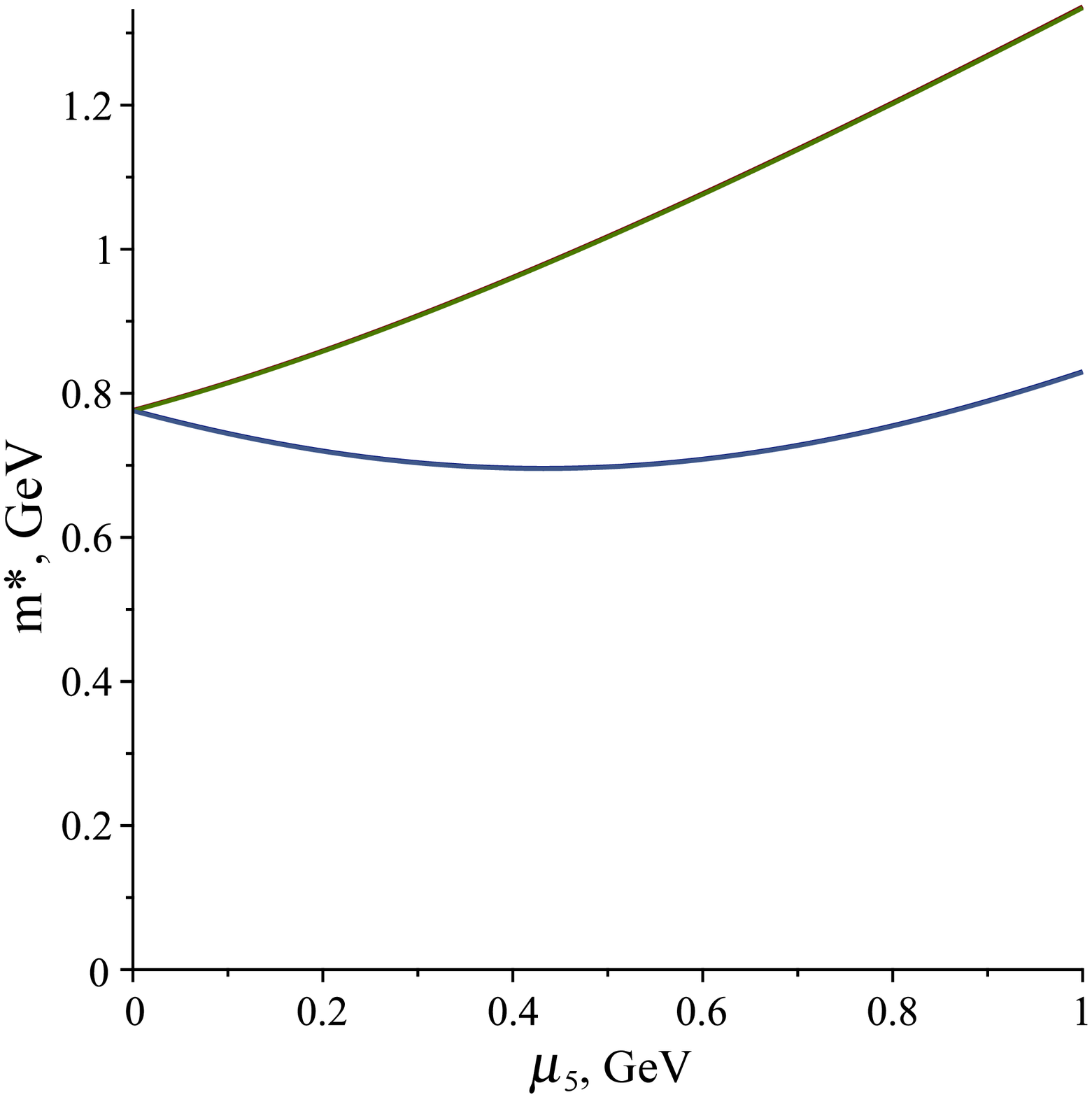} 
\vspace{-0.3cm}
\caption{Vector meson spectrum as a function of the chiral chemical potential at meson momentum $|\vec{k}|$=0.1 GeV and $|\vec{k}|$=0.4 GeV}\label{fig2}
\vspace{-0.3cm}
\end{figure}

We also plot the masses as a function of $\mu_5$ at $|\vec{k}|$=0.1 GeV and $|\vec{k}|$=0.4 GeV (figure \ref{fig2}).

We see that the massive fermion term $\xi b_0^2 |\vec{k}|^2/m^2$ enters the game at $\mu_5 k \geq$ 0.1 GeV$^2$. Its contribution is higher than the electromagnetic correction.

\section{Conclusions}
We calculated vector mesons spectrum in a medium with a chiral imbalance induced by the vacuum of fermions. We obtain
 additional, radiatively induced contribution that becomes important at momentum and $\mu_5$ around a few hundred MeV.

\end{document}